# The Pleiades: the celestial herd of ancient timekeepers.

Amelia Sparavigna
Dipartimento di Fisica, Politecnico di Torino
C.so Duca degli Abruzzi 24, Torino, Italy

**Abstract**
In the ancient Egypt seven goddesses, represented by seven cows, composed the celestial herd that provides the nourishment to her worshippers. This herd is observed in the sky as a group of stars, the Pleiades, close to Aldebaran, the main star in the Taurus constellation. For many ancient populations, Pleiades were relevant stars and their rising was marked as a special time of the year. In this paper, we will discuss the presence of these stars in ancient cultures. Moreover, we will report some results of archeoastronomy on the role for timekeeping of these stars, results which show that for hunter-gatherers at Palaeolithic times, they were linked to the seasonal cycles of aurochs.

**1. Introduction**
Archeoastronomy studies astronomical practices and related mythologies of the ancient cultures, to understand how past peoples observed and used the celestial phenomena and what was the role played by the sky in their cultures. This discipline is then a branch of the cultural astronomy, an interdisciplinary field that relates astronomical phenomena to current and ancient cultures. It must then be distinguished from the history of astronomy, because astronomy is a culturally specific concept and ancient peoples may have been related to the sky in different way [1,2].
Archeoastronomy is considered as a quite new interdisciplinary science, rooted in the Stonehenge studies of 1960s by the astronomer Gerald Hawkins, who tested Stonehenge alignments by computer, and concluded that these stones marked key dates in the megalithic calendar [3]. From that discovery, the study of how the ancient timekeepers used sun, moon and stars to subdivide the year strongly increased.
A group of stars, the Pleiades, shone in the sky as an important time-marker. The rising, heliacal or acronychal, of these stars announced to ancient populations a special period of the year or the starting of a new season. Pleiades and Aldebaran, the main red star in Taurus constellation, can be considered among the oldest objects in the sky observed by man. Their relevance is testified by objects and painting since Palaeolithic times. Let us see how mythology explains the origin of the cluster, starting with Greek mythology from which they gained the current name in astronomy.

**2. The Pleiades**
Also known as the Seven Sisters, the Pleiades are an open cluster in the constellation of Taurus [4,5]. These stars are among the nearest star clusters and the most obvious to the naked eye. Blue stars, observed through a faint nebulosity of a dust cloud in the interstellar medium, which the stars are currently passing through, dominate the cluster. The brightest stars of the cluster are named for the Seven Sisters of Greek mythology, daughters of Atlas and Pleione. Pursued by Orion, they were rescued by Zeus, who immortalised the sisters by placing them in the sky.

Due to a high visibility, these stars gained a special place in many ancient cultures. They are winter stars in the Northern Hemisphere and summer stars in the Southern Hemisphere: we can tell that these stars were known since old times, by several cultures all around the world, including the Maori and Australian Aborigines, Chinese, Maya and Aztec and the Native people of North America. The Pleiades are particularly important in Hindu mythology as the six wives of the six sages. The number is not fixed but changing in the myths between six and seven.

Of course, representations of these stars in the local mythologies are different, but a rather common element is their female nature. For instance, in one of the Maori traditions, Matariki, the Maori name for the cluster of stars, is a mother with her six daughters. The Sioux of North America had a legend linking the origin of Pleiades to the Devil's Tower. The stars were seven women, pursued by a bear. They prayed the gods, who raised the ground where they were located high into the air, to save them from the bear. The maidens then turned into stars.

The Pleiades are mentioned in Homer's Iliad and Odyssey [6]. In the Greek myths, several of Olympian gods were engaged with the seven heavenly sisters. Merope, the youngest of the seven Pleiades, married Sisyphus and, becoming mortal, faded away: this is how the myth explains why in the Pleiades star cluster only six of the stars shine brightly and the seventh, Merope, shines dully. The Pleiades start to shine over the horizon and set in the West, during October-November, the proper time of the year in Mediterranean area, to plough and sow the land. As prominent stars in the Greek agricultural calendar, Hesiod, the poet who lived around 700 BC, mentions the Pleiades several times in his poem, the Works and Days.

**3. Heliacal and acronychal rising.**

The heliacal rising of a star or of a constellation occurs when it becomes visible above the eastern horizon at dawn, after a period when it was hidden below the horizon. The corresponding rising of a celestial body above the eastern horizon at nightfall is its acronychal rising. Constellations that rise and set were then used for calendars by ancient peoples and used to form a zodiac. As highly visible heliacal stars, the Pleiades were among the most important celestial body, after the moon, and used for a first astronomic conception. The Pleiades heliacal rising was widely recognised in Austral regions, as the beginning of the new-year and then of agricultural season. To the Maori of New Zealand, their heliacal rising signifies the beginning of a new-year.

For the Bronze Age people of Europe, the Pleiades started to be associated with mourning and funerals. Between the autumn equinox and the winter solstice, the cluster rose in the eastern sky after the sunset: for Celtic population, these stars were a window into the Otherworld. Not surprisingly, this rising was coincident with a festival devoted to the remembrance of dead people [7]. As a result of precession over the centuries, the Pleiades no longer marked the festival, but the association nevertheless persists.

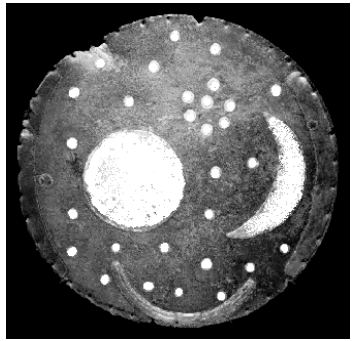

Fig.1 The Nebra Disk is usually considered one of the oldest representations of Pleiades.

An artefact (Fig.1), dated 1600 BC, the bronze disk from Nebra in Germany, is usually considered as one of the oldest known representations of the sky, where Pleiades are top right. This disk is of the Bronze Age: as we shall see in the next section, there are much older paintings depicting the Pleiades in the sky.

**4. The Palaeolithic timekeepers.**
It is necessary to recognise this fact: we have representations of stars of considerably earlier epochs. The Archeoastronomy confirmed that the Palaeolithic man was able to follow the prominent patterns of stars. In the Palaeolithic epochs, which lapsed from 33,000 to 10,000 BP (before present), the ancient cultures were able to develop calendars, as demonstrated by marks on transportable bones and stones and paintings on walls of the caves [8]. M.A. Rappenglück showed, that man also recognized, among the others, the Pleiades, as it is shown by paintings in the cave of Lascaux (France). The constellations were used by the Palaeolithic hunter-gatherers for orientation in space and for time reckoning. Of course, the patterns of stars in the sky assumed an important role in the spiritual life.

Rappenglück shows in [8], a panel in the cave of La-Tête-du-Lion (France) with a combination of Aldebaran in the Bull and the Pleiades. This picture (ca.21,000–22,000BP) has a remarkable similarity with the representation in the Lascaux cave (Fig.2). A female bovine, probably an aurochs-cow (Bos primigenius) is depicted in red ochre colour. Three groups of dots are related to this animal. One is a group of seven points, situated over the body. Note that the bovine image is representing a female, not a male bovine. Aurochs with the six spots may represent the Taurus constellation with Pleiades above.

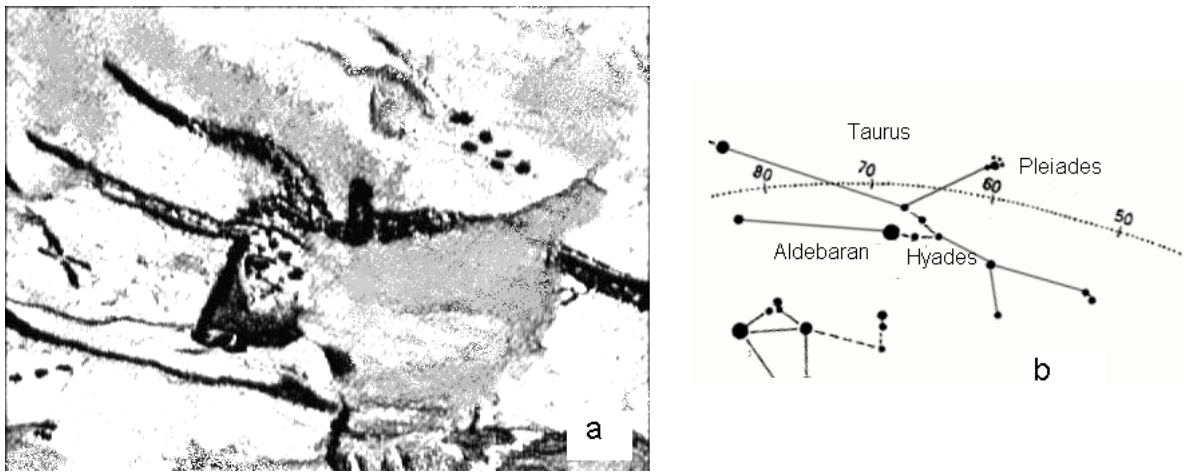

Fig.2 An aurochs on a wall of Lascaux cave (a) compared with the map of the sky (b), adapted from Ref.4.

The eye of the bovine marks Aldebaran (the name means "he who follows" in Arabic, since the star follows the Pleiades in its apparent motion) and the cluster of dots in the face of the animal is related to the cluster of Hyades. The cow image in La-Tête-du-Lion cave has a prominent dot, which marks Aldebaran and seven points representing the Pleiades. To prove the hypothesis, the author compares this image with depictions of the Pleiades due to different ancient cultures (Chinese and Navajo).

The Pleiades phases, depending on epochs and moon position, permit to determine the starting points of the lunar and solar years, and the subdivision of these years in shorter time intervals. This leads to fractionating the ecliptic into "houses". At around 2,300 BC, the Pleiades were the starting point of a "house", corresponding to the spring equinox, for ancient Chinese and Indians. The second "house" of the Indian lunar zodiac is referring to a red cow, Aldebaran [8]: it may be a reminiscence of the Palaeolithic "aurochs" constellation. The ancient Chinese, approximately from 2,300 BC, started to celebrate the full moon passing the Pleiades at the autumnal equinox.

This tradition is an excellent illustration of similar rites, which we can imagine as being performed by the hunter-gatherers at Palaeolithic times, linked to the life-cycles of aurochs. To support this idea, Rappenglück mentioned the traditions of some Natives of North America, who connected the sidereal year of Pleiades with the life-cycles of bison. To reinforce the Rappenglück's idea, we can add the fact that [9,10], Turkic and Mongol peoples related the beginning of seasons to the appearance and disappearance of the Pleiades at the horizon, devoting a number of rituals to such events. As nomad cattlemen, the Kazakhs co-ordinated their activities, mainly consisting in cattle breeding and hunting, with seasons of the year. Urker (the Pleiades) has a predominating position of among other heavenly bodies in their calendar.

**5. The Egyptian Seasons and the Seven Hathors.**
The ancient Egyptians based their calendar on the heliacal rising of Sirius and devised a method of telling time during the night, method based on the heliacal risings of 36 stars, a star for each 10-degrees segment of the 360-degrees zodiac circle [8]. The first of three seasons of the ancient Egyptian calendar is the inundation season. This was the time of the Egyptian calendar year when the Nile waters flooded the farmland [11]. The last month of this season is Ahtyr: this name is a variant of Hathor, the goddess guardian of the tombs. At Plutarch's time, Athyr month was coincident with October-November. As for Celtic population, we have a connection with the Otherworld.

Known today by the Greek name, Hathor is the Egyptian patroness of lovers, the goddess of the sky, the protector of women and children, and beloved of both the living and the dead (Fig.3). Earliest references of this goddess date back to the second dynasty. In art, she was often depicted with just the head or the whole body of a cow, the Heavenly Cow. Worshipped at the city of the dead, at Thebes, she became the Goddess of the Dead.

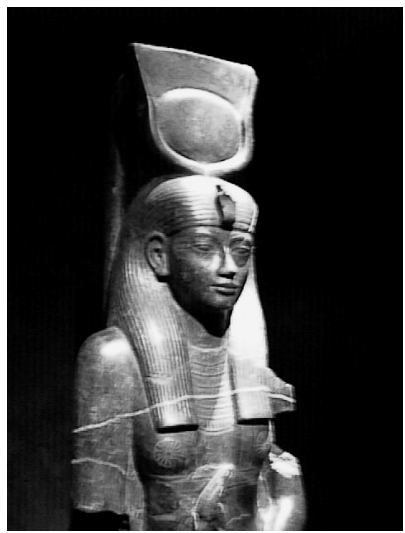

Fig.3 In Egyptian mythology, Hathor, in her form as the celestial cow, provides the sustenance to her worshippers. In earlier myths, Hathor was responsible for the raising of the Sun to the sky with her horns (Hathor statue at the Egyptian Museum of Torino).

The name Hathor means the "house of Horus" in the zodiac (the Heavenly Cow). It was during the Old Kingdom that she assumed the properties of an earlier bovine goddess, Bat [12]. She is also worshipped in the form of "Seven Hathors": these seven goddesses are the Pleiades shining in the sky [13], usually represented by seven cows, often associated with a bull, as a heaven herd providing the nourishment, bread and beer, in the Underworld [14]. We find again the Taurus, with Aldebaran its main red star, as one of the most ancient group of stars viewed as a constellation, also in Egyptian area. As the Seven Hathors, she was the goddess often present at birth. Able to foretell the future, she was connected with the Nile inundation and the abundance of the grain harvest [15].

The Seven Hathors of the Celestial Herd were named in a spell of the Book of the Dead and these names are: the "Lady of the Universe", the "Sky-storm", "The hidden one, presiding over her place", "You, from Khemmis", the "Red-hair", the "Bright Red" and "Your Name prevails over the West" [16]. Often accompanied by Osiris-Apis, Bull of the West, and the oars representing the four cardinal points, in the vignettes enclosed to the text in the "Book of the Dead", the seven cows and the bull are depicted in front of the offering tables of worshippers (Fig.4). The representation of animals (wild or domesticated) in front of an offering pole or table is also common in the Indus Valley civilization [17].

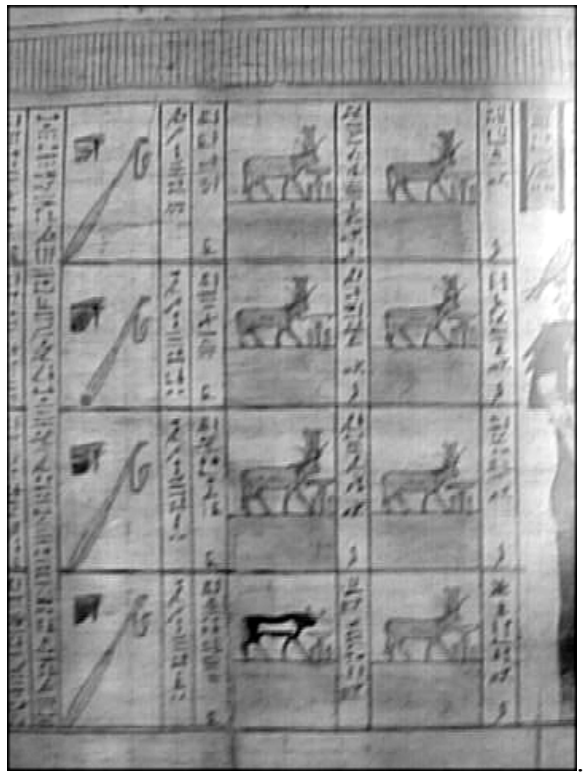

Fig.4. A vignette in the Egyptian Book of the Dead at the Egyptian Museum of Torino. On the left of the herd, the four oars, which represent the four cardinal points.

**6. Discussion.**
As we have seen, a highly probable cult of the wild cattle can be imagined at Palaeolithic times, connected with the Pleiades motion in the sky in the Mediterranean region. In Egypt, the cattle cult can be traced, from the worship of the Bos primigenius, the aurochs, till that of the domesticated cattle. And then, to the worship of Hathor. A detailed report on the subject is Ref.18.

The fact that Seven Hathors represent the Pleiades is not completely accepted [15]. Mainly this is due to the fact that it is easy to see just six stars. And the ancient people measured the keenness of vision by the number of stars the viewer could see in the Pleiades; in fact, it is six the number of stars usually seen, not seven. This explains why, in other cultures, for instance in the Indus Valley, the number of the Pleiades is six. Assuming the stars have not changed brightness significantly since ancient times, it is possible to guess that the reason is in the preference of the number seven [19]: the Hyades too were supposed to be seven in number, an abstraction even less plausible than that for the Pleiades.

The Greek mythology explains this mismatch telling that one of the stars is not easy to see because of her shame for being in love with a mortal. And the Egyptians tell that one Hathor is "The hidden one, presiding over her place": as a consequence in some paintings of the celestial herd, for instance in the Maihirpre's Tomb and in the Neferati's Tomb, one of the cows has a black coat. This is why we cannot see her.

**References.**
[1] R.M. Sinclair, (2005). The Nature of Archaeoastronomy, in J.W. Fountain and R.M. Sinclair: Current Studies in Archaeoastronomy, ISBN0890897719.
[2] C.L.N. Ruggles, (2005). Ancient Astronomy. ABC-Clio. ISBN1851094776.
[3] G.S. Hawkins and J.B. White, (1965), Stonehenge Decoded, Doubleday.
[4] Pleiades Observing Project, http://www.ast.cam.ac.uk/~ipswich/Observations/ Pleiades_Observing_Proj/POP.htm
[5] J. Herrmann, (1975) Atlante di astronomia, Oscar Mondadori, Milano.
[6] G. Ruggieri, (1967) Le meraviglie del cielo, Arnoldo Mondadori, Milano.
[7] A. Murphy and R. Moore, (2008), Island of the Setting Sun: In Search of Ireland's Ancient Astronomers, Liffey Press, www.mythicalireland.com
[8] M.A. Rappenglück, (1999) Earth, Moon, and Planets, Volumes 85-86, pp. 391-404(14).
[9] Nyssanbay Bekbassar, Pleiades in the Kazakh Ethnoastronomy, 22–31 July 2007, Klaipėda, Lithuania, 15[th] Annual Meeting of the European Society for Astronomy in Culture.
[10] Nyssanbay Bekbassar Astronomy in Kazakh Folk Culture, SEAC 2002 Tenth Annual Conference, 27–30 August, Tartu, Estonia.
[11] G. Hart, (2005) The Routledge Dictionary of Egyptian Gods and Goddesses, Routledge.
[12] http://en.wikipedia.org/wiki/Bat_(goddess)
[13] W. Max Muller, (2004). Egyptian Mythology, Kessinger Publishing.
[14] A. Feyerick, C.H. Gordon and N,M. Sharma, (1996). Genesis: World of Myths and Patriarchs, NYU Press, ISBN0814726682.
[15] G. Massey, (1998). The Natural Genesis, Black Classic Press, ISBN1574780093.
[16] Advanced Papyrological Information System, APIS record: chicago.apis.5429
[17] A. Sparavigna, (2008). Icons and signs from the ancient Harappa, CogPrints.org/6179
[18] M. Brass, (2002) Tracing the origins of the ancient Egyptian Cattle cult, at www.scribd.com/doc/1037932/Tracing-the-Origins-of-the-Ancient-Egyptian-Cattle-Cult
[19] S.J. Gibson, http://www.naic.edu/~gibson/